\documentclass[aps,pra,twocolumn,nofootinbib,groupedaddress,floatfix]{revtex4-2}

\usepackage{graphicx}
\usepackage{amssymb}
\usepackage{amsmath}
\usepackage{amsfonts}
\usepackage{multirow}
\usepackage{longtable}

\begin{document}

\title{
Enhancement of the molecular electron chirality by electronic excitation
}

\author{Naoya Kuroda}
\affiliation{Department of Micro Engineering, Kyoto University, Kyoto 615-8540, Japan}
\author{Masato Senami}
\email{senami@me.kyoto-u.ac.jp}
\affiliation{Department of Micro Engineering, Kyoto University, Kyoto 615-8540, Japan}

\date{\today}

\begin{abstract}

Previously, it was predicted that the parity-violating energy difference (PVED) between enantiomers 
and the molecular electron chirality (MEC),
which is the integrated value of the electron chirality density over the entire molecule
are enhanced for ionization or electronic excitation.
Following the confirmation of the enhancement of the PVED by electronic excitation in our previous papers,
the present work confirms that electronic excitation enhances 
the MEC for H$_2$Te$_2$ and H$_2$Se$_2$.
The PVED in the first excited state 
is explained by the contribution from the highest occupied molecular orbital
if the PVED contribution from the highest occupied molecular orbital in the ground state
dominates over any other contributions.
In this paper, 
it is checked whether
this explanation can also be applicable to the MEC.
It is also reported that the MEC is not the simple sum of contributions from each atom 
and 
the mixing by the superposition between atomic orbitals of different atoms is important.

\end{abstract}

\maketitle
\section{Introduction}

It is well established that certain molecules exhibit chiral structures,
while electrons also possess an intrinsic chirality,
i.e. the handedness of the electron can be defined.
A free electron does not have a specific charge distribution
and the handedness of the electron should not be defined by the distribution.
However, the electron possess spin,
and the inner product of spin and momentum
enables the definition of a specific handedness quantity known as helicity.
In the absence of external fields,
helicity is conserved over time,
since it commutes with the Hamiltonian operator.
It is noteworthy that 
helicity is frame-dependent and is not Lorentz invariant.
In contrast, the chirality of an electron is defined 
as a quantity that maintains a consistent handedness ratio across all inertial frames.
This concept represents a relativistic extension of helicity
as elucidated in the next section.

Nature inherently selects the chirality as the handedness quantity of electrons.
Notably, 
the weak interaction, one of the four fundamental interactions in particle physics,
differentiates between left- and right-handed electrons based on their chirality.
For instance, 
this interaction is mediated by W and Z gauge bosons,
and the W boson interacts with left-handed electrons
but does not with right-handed electrons.
Additionally,
in the framework of quantum field theory,
the torque density for the electron spin is 
predicted to be dependent on the electron chirality density
\cite{zetaforce, Tachibana:2017}.
In this theoretical prediction,
the local torque contribution, which is named the zeta force,
is expressed as the gradient of the electron chirality density.
While this contribution to torque density
has yet to be observed experimentally,
it has been corroborated by numerical simulations~\cite{confirmation}.

Although the chirality of an electron affects its physical properties,
free electrons eventually lose their chirality polarization due to mass-induced chirality flips.
For instance, if electrons are initially polarized predominantly in the left-handed chirality,
this polarization will decay over time to a state of no net polarization.
The left-handed component tends to convert to the right-handed component
more frequently than the reverse process occurs, 
as the initial ratio of left-handed to right-handed components is greater.

However, electrons within chiral molecules, which are parity-violating entities,
are expected to exhibit polarization in chirality.
This is because a state polarized in electron chirality 
may represent the most stable configuration in steady-state conditions due to spin-orbit interactions. 
Although the mass of the electron can induce chirality flips,
the ultimate objective is to achieve a stable polarized state in chirality.

One phenomenon associated with the electron chirality 
is the parity-violating energy difference (PVED) between enantiomers.
This energy difference arises from the Z-boson exchange interaction between nuclei and electrons
and is proportional to the difference in electron chirality density at the nuclei of the enantiomers.
Although extensive experimental efforts have yet to confirm this energy difference \cite{CHFClBr},
theoretical studies utilizing quantum chemistry computations 
indicate a nonzero PVED in many chiral molecules
\cite{Darquie:2010, Senami:2019, Kuroda:2022, Kuroda:2023}.
Importantly, this energy difference pertains solely to the electron chirality at the nuclei.
The electron chirality for the entire molecule of H$_2$Te$_2$ was reported to be nonzero
\cite{Bast:2011}.
Furthermore, systematic analyses in Ref.~\cite{Senami:2019, Senami:2018, Senami:2020}
demonstrate that chiral molecules generally exhibit a nonzero integrated value of electron chirality density
over the entire molecular structure. 
The existence of this chirality is named Electron Chirality in Chiral Molecules (ECCM) \cite{Senami:2020}.

The value of ECCM is small,
and phenomena related to ECCM are not extensively documented.
Nevertheless, our research group has investigated several associated phenomena.
One topic is the homochirality in nature \cite{text:homochirality}.
In an enantiomeric pair,
the number of the left-handed electron differs,
while the total number of electrons is the same.
Consequently, one enantiomer exhibits 
a higher reaction rate for weak interactions
than its counterpart.
This disparity leads to the preferential loss of one enantiomer through collisions
with weakly charged particles in both terrestrial and extraterrestrial environments \cite{Senami:2019}.
Another noteworthy phenomenon is 
Chirality-Induced Spin Selectivity (CISS) \cite{CISS}.
CISS is the strong spin polarization 
when electrons traverse chiral organic molecules such as DNA.
The mechanisms underlying this significant spin polarization remain unclear.
For this puzzle, 
examining the problem from the perspective of torque may provide critical insights, 
as opposed to focusing solely on spin-orbit interactions, which are typically framed in terms of energy. 
The torque experienced by electron spin is influenced by electron chirality density, 
with the torques generated by left- and right-handed electrons being oppositional.
ECCM predicts an opposite distribution of electron chirality within enantiomeric pair molecules,
suggesting that the zeta force may serve as the driving torque for CISS.
The correlation between spin polarization in CISS and the magnitude of the zeta force
was confirmed for alanine and helicene molecules \cite{Senami:2021}.

In the following discussion, 
the electron chirality of an entire molecule, 
which is the integrated value of the chirality density over the entire molecule,
is named Molecular Electron Chirality (MEC).
In Ref.~\cite{Senami:2019},
it was predicted that PVED and MEC are significantly enhanced 
for ionized or electronically excited states.\footnote{
In another viewpoint,
the enhancement of the PVED in electronic excited states
was discussed before our prediction \cite{Berger:2003}. }
In the ground state of H$_2$Te$_2$,
the contribution to PVED (and MEC) from a single valence orbital
exceeds the combined contributions from all other orbitals. 
This results in a relatively small PVED (MEC)
due to cancellations among the large contributions from different orbitals.
However, in ionized or excited states, such cancellations may be disrupted, 
potentially leading to substantial enhancements in both PVED and MEC compared to their values in the ground state.
This phenomenon of cancellation 
have been observed in various other molecules
\cite{Kuroda:2023,Wisenfeld:1988, Laerdahl:1999, Schwerdtfeger:2005}.
In Refs.~\cite{Kuroda:2022, Kuroda:2023},
the prediction regarding the enhancement in the PVED of
H$_2X_2$ ($X=$ O, S, Se, Te), CHFClBr, CHFClI, and CHFBrI has been confirmed.
In the study~\cite{Kuroda:2023},
the cancellation-breaking enhancement (CBE) hypothesis has been proposed,
suggesting that the PVED in the first electronic excited state 
is significantly greater than in the ground state
if the PVED contribution from the highest occupied molecular orbital (HOMO) in the ground state
surpasses the sum of contributions from all occupied orbitals.
Furthermore, 
if the HOMO contribution dominates over any other contribution,
the CBE hypothesis allows for the prediction of PVED values in the first excited state
based solely on the HOMO contribution.
For the enhancement of the PVED by the ionization, 
ionized states of CHFBrI derivatives such as CHDBrI$^+$ and CHCaBrI$^+$
are recently studied 
from the viewpoint of the experimental detection of parity-violation in vibrational transitions~\cite{Eduardus}.

In the present paper,
it is confirmed that
electronic excitation also leads to an enhancement of MEC.
Target molecules chosen for this investigation are
H$_2$Te$_2$ and H$_2$Se$_2$.
These H$_2$X$_2$ type molecules are often employed
to study parity-violating effects in molecules
\cite{Bast:2011,Laerdahl:1999,Berger:2005}.
In Refs.~\cite{Kuroda:2022, Kuroda:2023},
H$_2$O$_2$ and H$_2$S$_2$ are also studied for the enhancement of the PVED.
However, computing MEC for molecules that consist solely of light elements with small spin-orbit interactions,
such as H$_2$O$_2$ and H$_2$S$_2$
necessitates large basis sets and incurs significant computational costs \cite{Senami:2019,Senami:2020}.
Specifically, in Ref.~\cite{Senami:2019}, it was demonstrated that 
the dyall.ae3z basis set is insufficient for accurately calculating the MEC of H$_2$S$_2$,
while the dyall.ae4z basis set is required for reliable results.
Moreover, for the calculation of the MEC of H$_2$O$_2$,
even the dyall.ae4z basis set is insufficient.
The accurate calculation of H$_2$O$_2$ requires additional diffuse functions \cite{Senami:2020}. 
Consequently, H$_2$O$_2$ and H$_2$S$_2$ are excluded from the present study focused on excited states.
Nonetheless, we anticipate that 
H$_2$O$_2$ and H$_2$S$_2$ will exhibit results analogous to those of H$_2$Te$_2$ and H$_2$Se$_2$ 
due to the similar patterns of the total chirality in the ground state \cite{Senami:2019,Senami:2020}
and the enhancement of the PVED \cite{Kuroda:2022, Kuroda:2023}.
Following the confirmation of MEC enhancement,
it is studied whether the CBE hypothesis is applicable to the enhancement of the MEC.

This paper is organized as follows.
In the next section, 
the chirality of the electron is introduced,
and then
the computational method and details are explained in Sec.~\ref{sec:computation}.
In Sec.~\ref{sec:results}, 
our results of the enhancement of the MEC are shown.
First, the enhancement of the MEC by electronic excitation is confirmed 
and then the CBE hypothesis is applicable to the enhancement of the MEC.
The last section is devoted to the summary.

\section{Theory }
\label{sec:theory}

The MEC, the integrated value of the electron chirality density 
over the entire molecule,
is defined by the following formula,
\begin{align}
\int d^3 x \psi^\dagger (x) \gamma_5 \psi (x) ,
\end{align}
where $ \gamma_5 \equiv i \gamma^0 \gamma^1 \gamma^2 \gamma^3$
with the gamma matrices, $\gamma^\mu~(\mu=0-3)$
and $\psi$ is the electron wave function in a four-component representation.
The electron chirality density, $ \psi^\dagger (x) \gamma_5 \psi (x)$,
can be reduced to the combination form of 
the left- and right-handed electron components,
\begin{align}
\psi^\dagger (x) \gamma_5 \psi (x)
=
\psi_R^\dagger (x) \psi_R (x) 
- \psi_L^\dagger (x) \psi_L (x).
\end{align}
the left- and right-handed electron wave functions are 
defined as
\begin{align}
\psi_{R,L} (x) = P_{R,L} \psi (x),
\end{align}
with the projection operators,
\begin{align}
P_R = \frac{1 + \gamma_5}{2},
~~~ 
P_L = \frac{1 - \gamma_5}{2}.
\end{align}

The Lorentz transformation commutes with $\gamma_5$ \cite{textbook1}
and the chirality i.e. the ratio between the left- and right-handed electrons
are not changed in a different inertial frame.
From $\gamma_5$ operator, 
it is seen that the chirality is a relativistic extension of the helicity.
This operator is cast into another form,
which is the product of the velocity and spin operators,
\begin{align}
\gamma_5 
= \alpha_x \Sigma_x = \alpha_y \Sigma_y = \alpha_z \Sigma_z 
= \frac{1}{3} \sum_{i=1}^3 \alpha^i \Sigma^i .
\end{align}
Here, $\alpha^i = \gamma^0 \gamma^i$,
and $\Sigma^i = \gamma^0 \gamma^i \gamma_5 $,
which is the $4 \times 4 $ extension of the Pauli matrices.
Therefore, the latter operator is the spin operator.
The velocity operator is known to be defined as 
$\hat v^i = c \alpha^i $ with $\alpha^i$ matrices \cite{textbook1}.
Hence, $\gamma_5$ operator is the product of the velocity (divided by $c$) and spin operators
in relativistic quantum theory.
However, the observable of the chirality density is not 
the product of the observed velocity and spin, 
$ \psi^\dagger c \alpha^i \psi \psi^\dagger \Sigma^i \psi $,
and the observable of the chirality density is $\psi^\dagger \gamma_5 \psi$.
The former is not Lorentz invariant.

In this paper, the CBE hypothesis \cite{Kuroda:2023} is checked for the MEC.
In this paragraph, the CBE hypothesis is reviewed.
This hypothesis requires the following conditions:
the PVED is derived as the result of the cancellation between contributions from the HOMO and other orbitals,
the electron in the HOMO in the ground state is dominantly excited,
and other occupied orbitals in the ground state are not affected by the excitation.
These conditions are satisfied in the first electronic excitation state in many molecules \cite{Kuroda:2023}.
If these conditions are satisfied,
the PVED in the first excited state is much larger than the PVED in the ground state.
In addition, if the HOMO contribution to the PVED dominates over any other contributions,
the PVED value is roughly estimated only with the contribution from the HOMO,
\begin{align}
E_{\rm PV} ({\rm CBE})
=
- E_{\rm PV,HOMO} .
\label{eq:CBE_EPV}
\end{align}
Here $ E_{\rm PV,HOMO} $ is the contribution from one electron in the Kramers pair of the HOMO to $E_{\rm PV} $.
(Two electrons in the same Kramers pair give the same contribution to the electron chirality.)

While this CBE hypothesis is about the enhancement of the PVED,
the enhancement of the MEC is considered to occur in the same cancellation-breaking mechanism.
For the CBE hypothesis about the MEC, 
the conditions are the same as the PVED.
The estimate of the MEC is 
given as the HOMO contribution to the MEC,
that is,
\begin{align}
- \int d^3 x \psi_{\rm HOMO}^\dagger (x) \gamma_5 \psi_{\rm HOMO} (x) ,
\label{eq:CBE}
\end{align}
where $ \psi_{\rm HOMO} $ is the wave function of the HOMO.
In this paper, the CBE hypothesis is checked for the MEC.

\section{Computational detail }
\label{sec:computation}

The enhancement of the PVED of H$_2$Te$_2$ and H$_2$Se$_2$
was confirmed in our previous works \cite{Kuroda:2022, Kuroda:2023}.
In this work, 
the MEC of H$_2$Te$_2$ and H$_2$Se$_2$ in electronic excited states
is studied with quantum chemistry computations
and 
the enhancement of it is confirmed.
Then 
the enhancement mechanism of the MEC
is compared with that of the PVED by using these molecules.
The MEC of these molecules in the ground state
was studied in previous works \cite{Senami:2019, Senami:2020, Bast:2011}.

The structures of these molecules are chosen to be 
the same as our previous work \cite{Kuroda:2023}.
The difference in structure between ground and excited states 
is ignored for simplicity,
since the purpose of this paper is
the confirmation of the enhancement of the MEC by electronic excitation
and to study the mechanism of the enhancement.
The structures of H$_2$Te$_2$ and H$_2$Se$_2$ were determined 
by geometrical optimization computations 
with the DIRAC19 program package \cite{DIRAC19, DIRAC:2020}.
These computations were performed with the HF method with Dirac-Coulomb Hamiltonian
and the dyall.ae2z basis set \cite{ae2z, Dyall:2006}.
After the geometrical optimization,
the dihedral angle $\phi$ is taken as a free parameter
and the structure dependence of the enhancement is studied for the change of $ \phi $.

Computations of relativistic four-component wave functions 
are carried out with DIRAC19.
Computations of excited states are based on 
Equation-of-Motion Coupled-Cluster (EOM-CC) theory.
The MEC in excited states is calculated 
by using the Finite-Field Perturbation Theory (FFPT) \cite{Pople:1968, Pawlowski:2015} as explained later.
These computations are also supported by 
the comparison with results by the Z-vector method of Coupled Cluster Singles and Doubles (CCSD).
Electron chirality of a specific molecular orbital
is calculated with the HF method.
For HF computations,
the Dirac-Coulomb-Gaunt Hamiltonian is used,
and for CCSD and EOM-CCSD computations using the RELCCSD module \cite{Visscher:1995, Shee:2018},
molecular mean-field approximations to the Dirac-Coulomb-Gaunt Hamiltonian
are used \cite{Sikkema:2009}.
In all computations,
the Gaussian nuclei density distribution is used \cite{Visscher_Gauss:1997}.

In computations of H$_2$Te$_2$ and H$_2$Se$_2$, 
C$_2$ symmetry is adopted, and 
the dyall.acv3z basis set \cite{Dyall:2006, acvXz} is utilized.
In CCSD and EOM-CCSD computations of H$_2$Te$_2$ and H$_2$Se$_2$,
electrons in 4s4p4d (Te) and 3s3p3d (Se) orbitals are chosen for frozen core electrons.
The upper limit cutoff of the active space is 
70 Hartree (H$_2$Te$_2$) and 100 Hartree (H$_2$Se$_2$).
The threshold of the convergence in EOM-CCSD computations is $10^{-10}$. 

In our computations, 
the calculation of the MEC in excited states is done with the FFPT \cite{Pople:1968, Pawlowski:2015}.
The MEC is derived by
the perturbation calculation 
with the perturbation Hamiltonian $H_5 = \lambda \gamma_5 $,
where $\lambda $ is the perturbation parameter.
The expectation value of this operator is 
\begin{align}
\langle H_5 \rangle
=
\lambda \int d^3x \psi^\dagger (x) \gamma_5 \psi (x) 
=
\lambda \int d^3x \langle \gamma_5 \rangle .
\label{eq:H_P}
\end{align}
For this expectation value, 
the following relation is given with the Hellmann-Feynman theorem,
\begin{align}
\left. \frac{ \partial E \left( \lambda \right) }{\partial \lambda } \right|_{\lambda = 0}
&=
\left\langle \psi \left|
\frac{\partial H }{\partial \lambda } 
\right| \psi \right\rangle
= 
\int d^3x \psi^\dagger (x) \gamma_5 \psi (x) , 
\label{eq:HFT}
\end{align}
where $H$ is the total Hamiltonian
and $ \langle \psi |$ and $| \psi \rangle$ are bra and ket.
Using the finite difference method,
$\partial E / \partial \lambda |_{\lambda = 0} $ can be computed 
as 
\begin{align}
\left. \frac{\partial E }{ \partial \lambda } \right|_{\lambda = 0} 
\simeq 
\frac{ E (\lambda) - E(-\lambda) }{ 2\lambda } .
\end{align}
This computation of $ E (\pm \lambda) $ is done with the EOM-CCSD in the Dirac program.
For the finite difference method,
accurate computations require adequate value of $\lambda$.
For too large $\lambda$, effects from second or higher-order derivatives cannot be neglected,
while for too small $\lambda$, 
the effect of the perturbation is invisible,
since the effect is buried in a numerical error.
To search for an adequate value of $\lambda$,
the values of the MEC in the ground states are 
compared with those derived with CCSD computations.
Once the value of $ \lambda $ is determined, 
the same value of $ \lambda $ is used for excited states.
In the present work, 
$ \lambda =0.1$ is adopted for both H$_2$Te$_2$ and H$_2$Se$_2$
as explained later.


\section{Results}
\label{sec:results}

In this section, 
the atomic units are used for the electron chirality, particularly, the MEC.


\begin{figure}[tbp]
 \centering \includegraphics[width=0.9\linewidth]{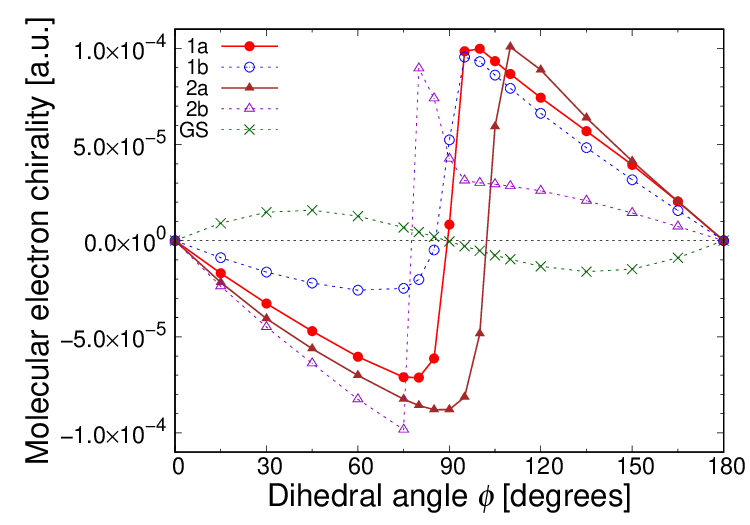}
\put(-120,-10){(a) H$_2$Te$_2$} 
\\
\vspace{5mm}
 \centering \includegraphics[width=0.9\linewidth]{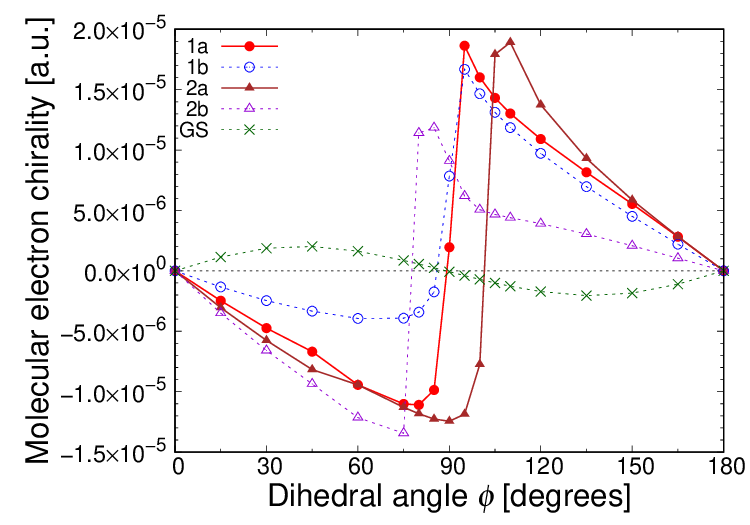}
\put(-120,-10){(b) H$_2$Se$_2$}
 \caption{
The molecular electron chirality in the ground and excited states
as a function of the dihedral angle.
The line named GS means the ground state.
$nx$ lines means the $n$-th excited state obeying the $x$-symmetry. 
}
 \label{fig:chirality_H2X2}
\end{figure}

Figure \ref{fig:chirality_H2X2} shows
the electron chirality in excited states
of (a) H$_2$Te$_2$ and (b) H$_2$Se$_2$ as a function of the dihedral angle.
The notation of excited states, $nx$, means the $n$-th excited state obeying the $x$-symmetry. 
The electron chirality in excited states
have opposite sign and much larger values
compared to that of the ground state (GS).
The enhancement of the MEC as well as the PVED
is confirmed.
While the MEC dependence on the dihedral angle in the ground state 
is a sigmoidal curve,
in excited states a different pattern of the dependence is seen.
All curves of the MEC have similar distributions.
In their patterns,
the MEC increases to 90$^\circ$
and flips the sign around $\phi = 90^\circ$.
The structure at $\phi = 90^\circ$ is chiral,
while the MEC is zero for the GS and some excited states.
The reason why the PVED is zero around $\phi = 90^\circ$ in the ground state
is explained in terms of the parity violation operator $\sigma \cdot p$ and spin-orbit interaction
using perturbation theory in Ref.~\cite{Faglioni:2001}.
Since the MEC and PVED has the same dependence on $\phi$,
the same discussion will clarify why the MEC is zero around $\phi = 90^\circ$ in the ground state.
Most excited states are derived as the excitation of one electron in the HOMO in the ground state.
The excitation ratio of the HOMO is roughly 90\%
except for 2a and 2b states around $\phi = 90^\circ$ for both molecules,
where the ratio is 20-70\%.
The energies of these excited states were reported in Ref.~\cite{Kuroda:2023}.
For both molecules, 
1a, 1b, and 2a ($\phi < 90^\circ$) or 1a, 1b, and 2b ($\phi > 90^\circ$)
are in the same approximate triplet state.
The rest 2a (2b) state is the singlet state for $\phi > 105^\circ$ $(\phi<75^\circ)$,
and forms approximate triplet state with 3a and 3b for $90^\circ < \phi < 105^\circ$ $(75^\circ < \phi < 90^\circ)$.

The most stable structure of these excited states of H$_2$Te$_2$ and H$_2$Se$_2$
are achiral ($\phi = 0^\circ $ or $ 180^\circ $).
Actually, the MEC for the optimized structure in these excited states
is zero, indicating no enhancement for these molecules in these excited states.
This study of H$_2$Te$_2$ and H$_2$Se$_2$ is
not for the search for molecules with large MEC in excited states
but for the confirmation of the enhancement in electronic excited states.
Moreover, the optimized structure of electronic excited states 3a, 3b, and 4a 
have $\phi = 90^\circ$ (3a and 3b) or $105^\circ$ (4a)~\cite{Kuroda:2023}.
Hence, some higher electronic excited states have chiral structures.

\begin{table}[tbp]
\caption{
Dependence of the MEC on $\lambda$ within the FFPT method
 in the electronic ground state of H$_2$Te$_2$ at $\phi = 45^\circ$.
 Dev. represents the relative deviation of the FFPT result from the Z-vector one.
}
\centering
\scalebox{1.0}{
\begin{tabular}{cccr}
\hline \hline
\begin{tabular}{c} Method \end{tabular} &
\begin{tabular}{c} $\lambda$ {[}a.u.{]} \end{tabular} &
\begin{tabular}{c} $\int d^3x \langle \gamma_5 (x) \rangle/10^{-5}$ {[}a.u.{]} \end{tabular} &
\begin{tabular}{c} Dev. {[}\%{]} \end{tabular} \tabularnewline
\hline
\multicolumn{1}{c}{FFPT} & $1.0\times10^{+1}$ & $0.581$ & $-64.2$\tabularnewline
 & $1.0\times10^{-0}$ & $1.574$ & $-2.8$\tabularnewline
  & $1.0\times10^{-1}$ & $1.585$ & $-2.2$\tabularnewline
 & $1.0\times10^{-2}$ & $1.558$ & $-3.9$\tabularnewline
 & $1.0\times10^{-3}$ & $1.440$ & $-11.1$\tabularnewline
\hline
Z-vector & - & $1.621$ & - \tabularnewline
\hline \hline
\end{tabular}
}
\label{tab:lambda_compare_GS_H2Te2}
\caption{Dependence of the MEC on $\lambda$ 
within the FFPT method in the electronic ground state of H$_2$Se$_2$ at $\phi = 45^\circ$.
 Dev. represents the relative deviation of the FFPT result from the Z-vector one.
}
\centering
\scalebox{1.0}{
\begin{tabular}{cccr}
\hline \hline
\begin{tabular}{c} Method \end{tabular} &
\begin{tabular}{c} $\lambda$ {[}a.u.{]} \end{tabular} &
\begin{tabular}{c} $\int d^3x \langle \gamma_5 (x) \rangle/10^{-6}$ {[}a.u.{]} \end{tabular} &
\begin{tabular}{c} Dev. {[}\%{]} \end{tabular} \tabularnewline
\hline
\multicolumn{1}{c}{FFPT} & $1.0\times10^{+1}$ & $-0.323$ & $-115.4$\tabularnewline
 & $1.0\times10^{-0}$ & $1.998$ & $-4.5$\tabularnewline
  & $1.0\times10^{-1}$ & $2.025$ & $-3.3$\tabularnewline
  & $1.0\times10^{-2}$ & $1.984$ & $-5.2$\tabularnewline
 & $1.0\times10^{-3}$ & $1.985$ & $-5.2$\tabularnewline
\hline
Z-vector & - & $2.093$ & - \tabularnewline
\hline \hline
\end{tabular}
}
\label{tab:lambda_compare_GS_H2Se2}
\end{table}

As explained in the previous section,
the usage of the adequate value of $\lambda$
is important for the evaluation of the MEC with the FFPT method.
In our work, the value of $\lambda$ is 
determined by the comparison with the value of the MEC
in the CCSD computation using the Z-vector method.
The dependence of the value of the MEC on $\lambda$ in the FFPT computations
is shown in 
Tables \ref{tab:lambda_compare_GS_H2Te2} and \ref{tab:lambda_compare_GS_H2Se2}.
For these computations, the dihedral angle is chosen to be $\phi = 45^\circ$,
since the integrated chirality is almost zero in optimized structure $(\phi = 90^\circ$)
and large at $\phi = 45^\circ$.
The deviation is the smallest at $\lambda = 0.1$,
and this value is adopted for both molecules.

\begin{table}[tbp]
\caption{
Dependence of the MEC on $\lambda$ 
within the FFPT method in excited states of H$_2$Te$_2$
at $\phi = 45^\circ$.
}
\centering
\scalebox{1.0}{
\begin{tabular}{ccr}
\hline
\hline
\begin{tabular}{c}  $\lambda$ {[}a.u.{]} \end{tabular} &
\begin{tabular}{c} State \end{tabular} &
\begin{tabular}{c} $\int d^3x \langle \gamma_5 (x) \rangle/10^{-5}$ {[}a.u.{]} \end{tabular}  \tabularnewline
\hline
$1.0\times10^{-0}$     
                               &1a       &$-4.656  $  \tabularnewline
                               &1b      &$ -2.182 $  \tabularnewline
                               & 2a   &$-5.567  $  \tabularnewline
                               &2b    &$ -6.325 $  \tabularnewline
\hline
$1.0\times10^{-1}$    
                               &1a  &$ -4.707 $  \tabularnewline
                               &1b   &$ -2.205 $  \tabularnewline
                               & 2a  &$ -5.626 $  \tabularnewline
                               & 2b  &$ -6.389 $  \tabularnewline
\hline
$1.0\times10^{-2}$   
                               &1a  &$ -4.744 $  \tabularnewline
                               &1b   &$ -2.240 $  \tabularnewline
                               & 2a  &$ -5.653 $  \tabularnewline
                               & 2b  &$ -6.425 $  \tabularnewline
\hline
\hline
\end{tabular}
}
\label{tab:lambda_compare_ES_H2Te2}
\end{table}

This single value of $ \lambda $ was used for all excited states and all angles of the same molecule
in the computation of Fig.~\ref{fig:chirality_H2X2}.
Even for excited states, 
the small change of $\lambda $ does not affect our results for excited states 
as seen in Table \ref{tab:lambda_compare_ES_H2Te2}.
The adequate value of $\lambda $ is 
determined by the ratio of the perturbation energy $ \int d^3x \langle \gamma_5 (x) \rangle $ to the molecular one.
The change of the molecular energy is small for the excitation and the change of the dihedral angle
compared to the increase of $ \int d^3x \langle \gamma_5 (x) \rangle $ by the excitation.
The increase of $ \int d^3x \langle \gamma_5 (x) \rangle $ from the ground state at $\phi = 45^\circ$
is several to ten times from Fig.~\ref{fig:chirality_H2X2}.
In Tables \ref{tab:lambda_compare_GS_H2Te2} and \ref{tab:lambda_compare_GS_H2Se2},
the deviation is less than 3\% (H$_2$Te$_2$) and 5\% (H$_2$Se$_2$) for $\lambda = 1.0 $
and hence
for ten times larger perturbation Hamiltonian $\lambda \int d^3x \langle \gamma_5 (x) \rangle$
in the ground state at $\phi = 45^\circ$ 
the deviation is confirmed to be acceptably small.
Hence, 
the usage of the same $\lambda $ in excited states is considered to be appropriate.

\begin{figure}[t]
 \centering \includegraphics[width=0.9\linewidth]{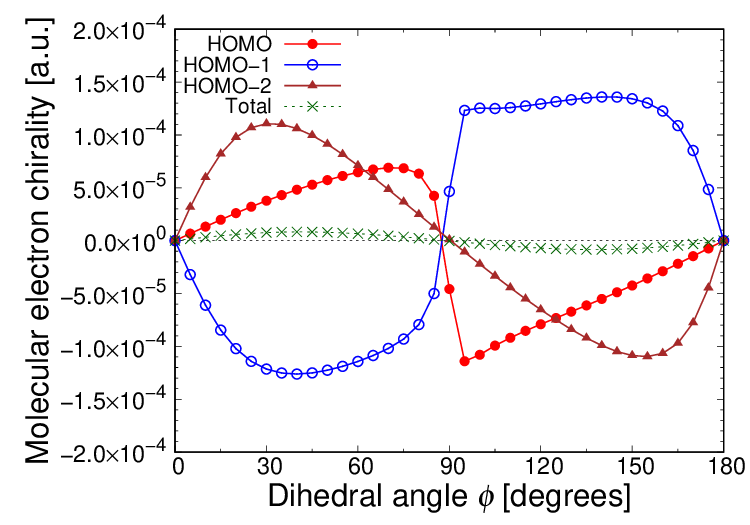}
\put(-120,-10){(a) H$_2$Te$_2$}
\\
 \vspace{20pt}
 \centering \includegraphics[width=0.9\linewidth]{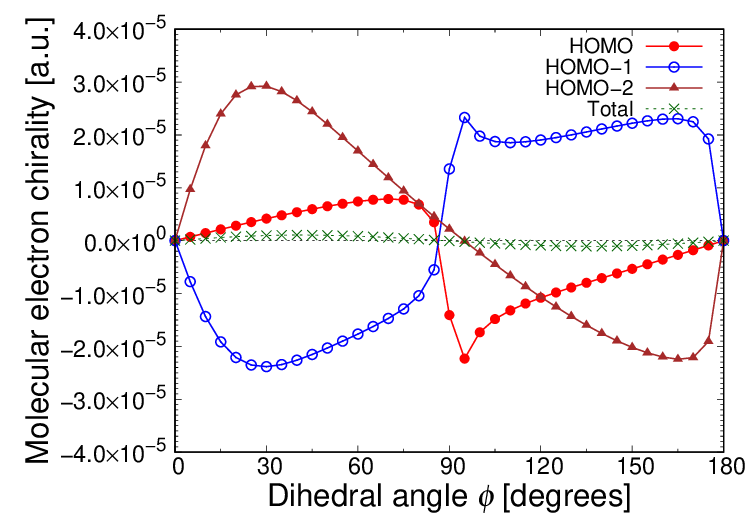}
\put(-120,-10){(b) H$_2$Se$_2$}
 \caption{
Contributions from single molecular orbitals to the MEC
as a function of the dihedral angle
as well as the total electron chirality summed up all the occupied orbitals.
}
 \label{fig:orbitals_H2X2}
\end{figure}

To understand the dependence of the MEC on the dihedral angle,
contributions from single molecular orbitals are studied.
In Fig.~\ref{fig:orbitals_H2X2},
contributions from HOMO, HOMO-1, and HOMO-2
are shown as a function of the dihedral angle.
For both molecules,
contributions from HOMO, HOMO-1, and HOMO-2
are much larger than the total electron chirality summed up all the occupied orbitals.
These three contributions are almost canceled out.
This is the key to the idea of the enhancement of the electron chirality
and is consistent with the previous report \cite{Senami:2019}.
The pattern of the HOMO contribution is opposite to that of the MEC of 1a and 1b excited states.
This is consistent with the CBE hypothesis.

\begin{figure}[t]
 \centering \includegraphics[width=0.9\linewidth]{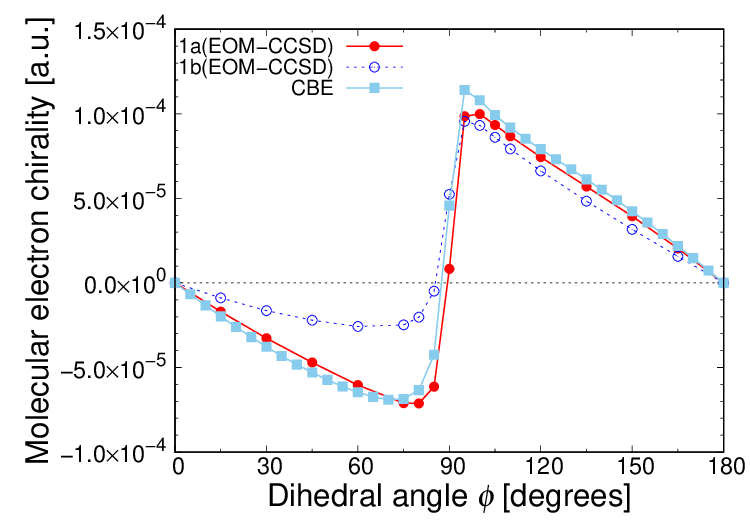}
\put(-120,-10){(a) H$_2$Te$_2$}
\\
 \vspace{20pt}
 \centering \includegraphics[width=0.9\linewidth]{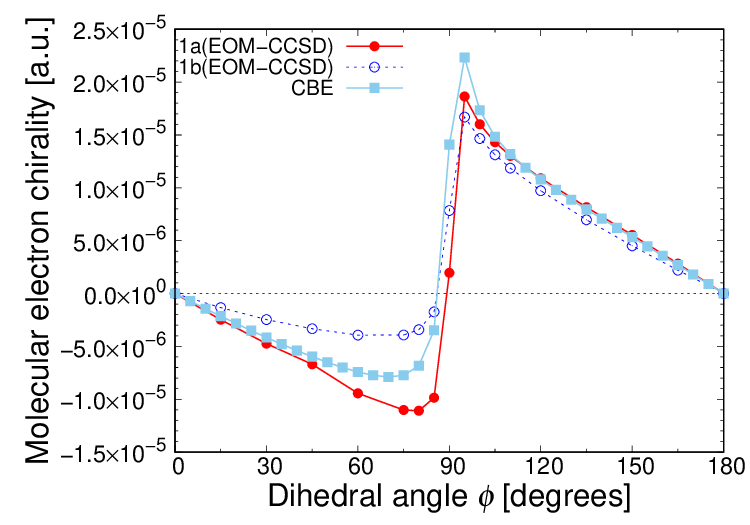}
\put(-120,-10){(b) H$_2$Se$_2$}
 \caption{
The prediction of the MEC by the CBE hypothesis.
}
 \label{fig:CBE_H2X2}
\end{figure}

In our previous paper \cite{Kuroda:2023},
it is reported that 
the value of the PVED in the first excited state of some molecules
can be predicted as the opposite sign of the HOMO contribution to the PVED
in the CBE hypothesis.
In this hypothesis,
when in the ground state the PVED contribution from the HOMO is larger than the total value of occupied orbitals,
the PVED in the first excited state is much larger than that in the ground state
by breaking the cancellation between contributions from the HOMO and the other orbitals.
Moreover, the PVED contribution from the HOMO in the ground state
dominates over any contribution from other occupied orbitals,
the PVED in the first excited state can be estimated from only the HOMO contribution.
Since the MEC of H$_2$Te$_2$ and H$_2$Se$_2$ is also derived as
a result of the cancellation between contributions from the HOMO and others orbitals~\cite{Senami:2019},
the conditions of the CBE hypothesis are satisfied for the MEC in H$_2$Te$_2$ and H$_2$Se$_2$.

In Fig.~\ref{fig:CBE_H2X2},
the estimate of the MEC in the CBE hypothesis 
is shown.
The predictions by the CBE hypothesis well realize 
the patterns of the MEC in 1a and 1b states.
Particularly, 
all predictions for $\phi > 90^\circ$
are perfectly consistent with computational results of the MEC.
For $\phi < 90^\circ$
the prediction in the 1a state of H$_2$Te$_2$
realizes precisely the MEC curve,
while  prediction curves for other states
have a few hundred percents deviations.
This inconsistency is considered to arise from 
the modification of occupied orbitals from the ground state
and the difference between HF and CCSD computations.
Nevertheless, 
the patterns of curves are consistent with the computational results.
The CBE hypothesis is considered to apply the MEC.

\begin{figure}[t]
 \centering \includegraphics[width=0.9\linewidth]{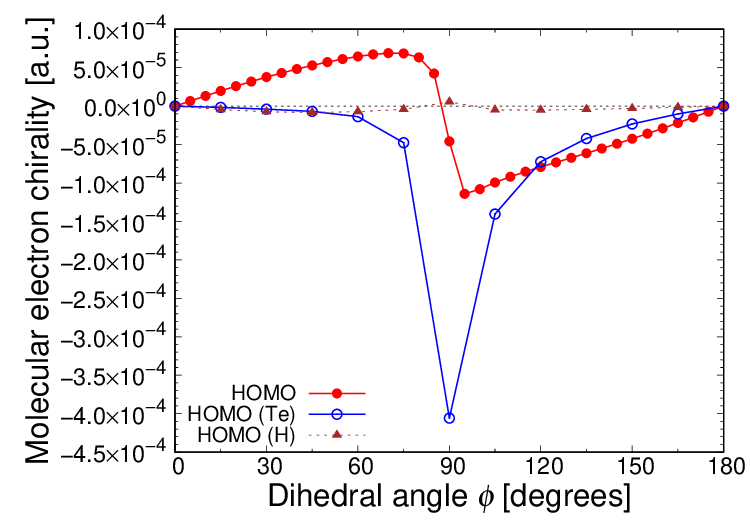}
\put(-120,-10){(a) H$_2$Te$_2$}
\\
 \vspace{20pt}
 \centering \includegraphics[width=0.9\linewidth]{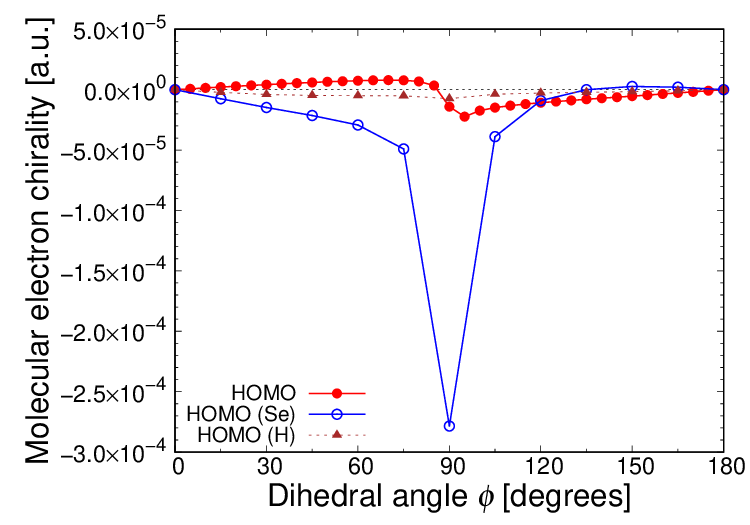}
\put(-120,-10){(b) H$_2$Se$_2$}
 \caption{
Atomic contribution in the HOMO to the MEC as a function of the dihedral angle.
For HOMO(Te, Se) and HOMO(H),
the HOMO contribution to the MEC is divided into atomic contributions,
where the contribution from atomic orbitals of only a single atom (Te, Se, or H)
is taken into consideration.
}
 \label{fig:Atom_H2X2}
\end{figure}

Finally, the dependence of the MEC on the dihedral angle is discussed.
It is known that the dependence of the MEC and PVED on the dihedral angle 
is almost the same in the ground state of H$_2$Te$_2$ and H$_2$Se$_2$ \cite{Senami:2019, Senami:2020, Bast:2011}.
For excited states, 
MEC curves have different patterns compared to PVED ones.
The pattern of MEC curves is like the teeth of a saw as shown in Fig.~\ref{fig:chirality_H2X2},
while PVED curves have a narrow peak~\cite{Kuroda:2023}.
The difference in the dependence on the dihedral angle between the MEC and the PVED in the first excited states 
can be elucidated by the CBE hypothesis,
and the dependence is almost explained by the HOMO contribution dependence on the dihedral angle
as seen from Eqs.~(\ref{eq:CBE_EPV}) and (\ref{eq:CBE}).
Therefore, 
the difference in the dependence is the difference in the HOMO contribution.
It is known that the PVED can be divided into contributions from each atom~\cite{Laerdahl:1999}.
Hence, to understand the difference in the HOMO contribution more deeply,
the HOMO contribution to the MEC is divided into atomic contributions,
i.e. heavier (Te or Se) and lighter (H) atomic contributions.
In Fig.~\ref{fig:Atom_H2X2},
atomic contributions in the HOMO to the MEC
is shown as a function of the dihedral angle.
For lines of HOMO(Te, Se) and HOMO(H),
the contribution from atomic orbitals of only a single atom (Te, Se, or H)
is taken into consideration.
These contributions are calculated with
QEDynamics (module for Dirac) \cite{QEDynamics}.
As seen from Fig.~\ref{fig:Atom_H2X2},
the HOMO contribution to the MEC 
is not the simple sum of atomic contributions
and the mixing by the superposition between atomic orbitals of heavier and lighter elements
is inevitably important.
This feature is a salient contrast to the PVED.
In Ref.~\cite{ Bast:2011},
it was implied that the MEC as well as the PVED is also the very atomic nature,
since the electron chirality is derived as the product of the small and large components
of the four-component wave function.
Our result shows that the MEC is not so simple and not atomic nature.
This is also the reason why
computations of the MEC for molecules with only light elements such as H$_2$O$_2$ 
requires additional diffuse functions for basis sets,
which reported in previous work~\cite{Senami:2020}.

\section{Summary}

For many molecules in the ground state,
the small value of the PVED and MEC
is derived as a result of the cancellation among larger contributions of valence electrons.
Hence, it was predicted that the PVED and MEC are enhanced 
for ionized or electronic excited states
by the cancellation breaking~\cite{Senami:2019}.
Our previous papers \cite{Kuroda:2022, Kuroda:2023}
showed that 
the PVED of H$_2X_2$ ($X=$ O, S, Se, Te), CHFClBr, CHFClI, and CHFBrI
are enhanced in electronic excited states.
In this work, it has been shown that 
electronic excitation really enhances 
the electron chirality of the entire molecule, MEC,
for H$_2$Te$_2$ and H$_2$Se$_2$ as well as the PVED.
In our previous papers \cite{Kuroda:2023},
the PVED in the first excited state 
is explained by the HOMO contribution to the PVED
if the PVED contribution from the HOMO in the ground state
dominates over any other contribution.
In this paper, 
it has been confirmed that 
this explanation can also be applicable to the MEC.
Then, the difference in the dependence on the dihedral angle between the MEC and PVED in excited states
has been discussed.
The dependence is determined by the dependence of HOMO contributions to the MEC and PVED.
While the PVED is known to be able to be divided into contributions from each atom,
it has been shown that the MEC is not explained with separate contributions from each atom.


\begin{acknowledgments}
This work was supported by Grants-in-Aid for Scientific Research (21H00072, and 22K12060).
The work of N. K. was supported by Grant-in-Aid for JSPS Fellows (23KJ1188)
and JST SPRING, Grant Number JPMJSP2110.
In this research work we used the supercomputer of ACCMS, Kyoto University.
The authors are grateful to Ayaki Sunaga for the contribution at the early stage of this work.
\end{acknowledgments}

\end{document}